\documentclass[a4paper]{jpconf}
\usepackage{graphicx}
\usepackage{amssymb,amsmath,amsfonts}
\usepackage{bm}
\usepackage{hyperref}
\usepackage{bm}

\newcommand{\dE}{I \! \! E}
\newcommand{\dB}{I \! \! B}

\newcommand{\dF}{I \! \! F}

\begin{document}

\title{Dynamics of classical and quantum spin in external fields: A general formalism}

\author{Yuri N. Obukhov}

\address{Theoretical Physics Laboratory, Nuclear Safety Institute, 
Russian Academy of Sciences, B.Tulskaya 52, 115191 Moscow, Russia}

\ead{obukhov@ibrae.ac.ru}

\begin{abstract}
We study the dynamics of classical and quantum particles with spin and dipole moments in external fields within the framework of the general approach by making use of the projection technique. Applications include the neutrino physics in matter, polarization phenomena in astrophysics, high-energy and heavy-ion physics. 
\end{abstract}

\section{Introduction}\label{intro}

Study of the dynamics of physical objects with spin (test particles and extended bodies, classical and quantum) in a curved space-time with arbitrary metric under the influence of arbitrary external fields represents a fundamental problem in theoretical physics. The knowledge of the general quantum-mechanical and classical equations of motion of a particle and spin is needed to test the foundations of fundamental physics, such as the relativistic invariance and possible violations of the Lorentz symmetry, the equivalence principle, the geometric structure of the space-time. Furthermore, this study is crucial for understanding of the behavior of astrophysical systems such as binary stars and is fundamentally important for the development of new physical devices (in particular, of new gravitational wave detectors), and for solving practical problems in application to to precision experiments in the high-energy physics and the heavy-ion physics.

In the framework of the Mathisson-Papapetrou-Dixon approach \cite{math,Dixon1}, the test particle is described by the 4-velocity $U^\alpha$ (normalized as  $U^\alpha U_\alpha = c^2$) and the tensor of spin $S^{\alpha\beta} = -\,S^{\beta\alpha}$. In general case, we assume that the particle has nontrivial magnetic and electric moments specified by 
\begin{equation}
\mu' = a\,{\frac {e\hbar}{2m}},\qquad \delta' = b\,{\frac {e\hbar}{2mc}}, \label{mude}
\end{equation}
where dimensionless constant parameters $a = (g -2)/2$ with the gyromagnetic factor $g$, and $b$ characterize the magnitude of the anomalous magnetic and electric dipole moments, respectively.

The equation of motion of spin under the action of the gravitational (inertial) and electromagnetic fields reads \cite{Dixon1,corben}
\begin{equation}\label{dotSab}
{\frac {DS^{\alpha\beta}}{d\tau}} - {\frac {U^\alpha U_\gamma}{c^2}}{\frac {DS^{\gamma\beta}}{d\tau}}
- {\frac {U^\beta U_\gamma}{c^2}}{\frac {DS^{\alpha\gamma}}{d\tau}}
= {}^\bot\!\Phi^\alpha{}_\gamma S^{\gamma\beta} - {}^\bot\!\Phi^\beta{}_\gamma S^{\gamma\alpha}. 
\end{equation}
Here $D/d\tau$ is the covariant derivative along particle's path parametrized by the proper time $\tau$, and $\Phi_{\alpha\beta} = - {\frac {e}{m}}\,F_{\alpha\beta} - {\frac 2\hbar}\,{}^\bot\!M_{\alpha\beta}$ with the generalized polarization tensor
\begin{equation}\label{Mab}
M_{\alpha\beta} = \mu' F_{\alpha\beta} + {\frac {c\delta'}2}\,\eta_{\alpha\beta\mu\nu}F^{\mu\nu}.
\end{equation}
Here $\eta_{\alpha\beta\mu\nu}$ is the totally antisymmetric Levi-Civita tensor. The projections ${}^\bot\!\Phi_{\alpha\beta} = \rho_\alpha^\mu\rho_\beta^\nu\Phi_{\mu\nu}$, ${}^\bot\!M_{\alpha\beta} = \rho_\alpha^\mu\rho_\beta^\nu M_{\mu\nu}$ are constructed with the help of $\rho_\alpha^\mu = \delta_\alpha^\mu - {\frac {1}{c^2}}U^\mu U_\alpha$. Provided the spin tensor satisfies Frenkel's supplementary condition $U_\alpha S^{\alpha\beta} = 0$, we can introduce the vector of spin
\begin{equation}\label{SV}
S^\alpha := \eta^{\alpha\beta\mu\nu}U_\beta S_{\mu\nu},\qquad 
S^{\alpha\beta} = {\frac 1{2c^2}}\,\eta^{\alpha\beta\mu\nu}U_\mu S_\nu,
\end{equation}
and then the equation of motion (\ref{dotSab}) is recast into
\begin{equation}
{\frac {DS^{\alpha}}{d\tau}} = \Phi^\alpha{}_\beta\,S^\beta.\label{dotSa}
\end{equation}
The physical meaning of (\ref{SV}) is that the spin has a purely ``magnetic'' nature. Namely, recall that in Maxwell's theory the magnetic and electric fields can be constructed in a covariant way as the projections of the field strength tensor $F_{\alpha\beta}$:
\begin{equation}
E_\alpha := F_{\alpha\beta}U^\beta,\qquad B^\alpha := {\frac 1{2c}}\eta^{\alpha\mu\nu\beta}
F_{\mu\nu}U_\beta.\label{EBmax}
\end{equation}
As a result, one has an unique representation of the electromagnetic field strength tensor in terms of its projections (electric and magnetic fields):
\begin{equation}
F_{\alpha\beta} = {\frac 1{c^2}}\left(E_\alpha U_\beta - E_\beta U_\alpha + c\eta_{\alpha\beta\mu\nu}
U^\mu B^\nu\right).\label{FEB}
\end{equation}

\section{General formalism for spin in arbitrary external fields}\label{pro}

Using observations above, we can develop a generalized model for dynamics of spin in arbitrary external fields by making use of the decomposition of the relativistic forces and moments of forces acting on a particle into ``magnetic'' and ``electric'' components. We will describe the motion of a test particle by the 4-velocity $U^\alpha$ and the 4-vector of spin $S^\alpha$, subject to the normalization $U_{\alpha} U^\alpha = c^2$ and the orthogonality condition $S_{\alpha} U^\alpha = 0$. In general, the dynamic equations for these variables read
\begin{eqnarray}
{\frac {dU^\alpha}{d\tau}} = {\cal F}^\alpha,\qquad
{\frac {dS^\alpha}{d\tau}} = \phi^\alpha{}_\beta S^\beta.\label{dotUS}
\end{eqnarray}
Here the forces ${\cal F}^\alpha$ on the particle, and the spin transport matrix $\phi^\alpha{}_\beta$ which affects the spin, may depend arbitrarily on external fields of different physical nature (electromagnetic, gravitational, scalar, etc.). The skew symmetry of the spin transport matrix, $\phi_{\alpha\beta} = - \phi_{\beta\alpha}$, guarantees the property $S_\alpha S^\alpha=$const. Normalization and orthogonality of the velocity and spin vectors impose conditions on the right-hand sides of the evolution system (\ref{dotUS}):
\begin{equation}
U_\alpha {\mathcal F}^\alpha = 0,\qquad U_\alpha\phi^\alpha{}_\beta S^\beta =
-\,S_\alpha {\mathcal F}^\alpha.\label{cc}
\end{equation}
Accordingly, the vector of the generalized force ${\mathcal F}^\alpha$ acting on a particle has only a transversal projection. As for the spin transport matrix $\phi_{\alpha\beta}$, like any bivector, we can decompose it into a pair of 4-vectors. Using (\ref{EBmax}) as a basic pattern, we define ``electric'' and ``magnetic'' projections
\begin{equation}
N_\alpha := \phi_{\alpha\beta}U^\beta,\qquad Q^\alpha := {\frac 1{2c}}\eta^{\alpha\mu\nu\beta}
\phi_{\mu\nu}U_\beta.\label{NQ}
\end{equation}
By construction, these 4-vectors are orthogonal to the velocity of the particle,
\begin{equation}
U_\alpha N^\alpha = 0,\qquad U_\alpha Q^\alpha = 0,\label{cc1}
\end{equation}
and the relations (\ref{cc}) are then recast into
$U_\alpha {\mathcal F}^\alpha = 0$ and $S_\alpha N^\alpha = S_\alpha {\mathcal F}^\alpha$. 
Therefore, we recover an unique decomposition analogous to (\ref{FEB}) in electromagnetism:
\begin{equation}
\phi_{\alpha\beta} = {\frac 1{c^2}}\left(N_\alpha U_\beta - N_\beta U_\alpha + c\eta_{\alpha\beta\mu\nu}
U^\mu Q^\nu\right).\label{PNQ}
\end{equation}
The orthogonality conditions (\ref{cc1}) can be resolved so that
\begin{equation}
N^0 = {\frac {1}{c^2}}(\bm{N}\cdot\widehat{\bm{v}}),\label{N0}
\end{equation}
where we use the boldface notation for 3-vectors: $\widehat{\bm{v}} = \{\widehat{v}^a\}$ and $\bm{N} = \{N^a\}$.

After these preliminaries, we are in a position to consider the dynamics of the physical spin which, as an ``internal angular momentum'' of a particle, is defined with respect to particle's rest frame system, in which $u^\alpha = \delta^\alpha_0$. The rest frame and the laboratory reference system are related via the Lorentz transformation $U^\alpha = \Lambda^\alpha{}_\beta u^\beta$. The latter reads explicitly
\begin{equation}
\Lambda^\alpha{}_\beta = \left(\begin{array}{c|c}\gamma & \gamma\widehat{v}{}_b/c^2 \\
\hline \gamma\widehat{v}{}^a & \delta^a_b + {\frac {\gamma - 1}
{\widehat{v}{}^2}}\,\widehat{v}{}^a\widehat{v}{}_b\end{array}\right),\label{Lam}
\end{equation}
where we parametrize the 4-velocity $U^\alpha$ in terms of the 3-velocity $\widehat{v}^a$ and the Lorentz factor $\gamma$:
\begin{equation}\label{U}
U^\alpha = \left(\begin{array}{c}\gamma \\ \gamma \widehat{v}^a\end{array}\right),\qquad
\gamma = {\frac {1}{\sqrt{1 - \widehat{v}^2/c^2}}},\qquad \widehat{v}^2 = \delta_{ab}\widehat{v}^a\widehat{v}^b.
\end{equation} 
The dynamic equation for the {\it physical spin} $s^\alpha = (\Lambda{}^{- 1})^\alpha{}_\beta S^\beta$ is directly derived from (\ref{dotUS}): 
\begin{equation}
{\frac {ds^\alpha}{d\tau}} = \Omega^\alpha{}_\beta s^\beta,\label{dsdt}
\end{equation}
where the tensor of angular precession of spin is constructed as
\begin{equation}\label{Omab1}
\Omega^\alpha{}_\beta = (\Lambda^{-1})^\alpha{}_\gamma\phi^\gamma{}_\delta\Lambda^\delta{}_\beta 
- (\Lambda^{-1})^\alpha{}_\gamma{\frac d {d\tau}} \Lambda^\gamma{}_\beta.
\end{equation}
In (\ref{dsdt}), the 0th component vanishes, which is equivalent to the second condition (\ref{cc}), and the spin evolution system (\ref{dsdt}) reduces to the 3-vector form
\begin{equation}
{\frac {ds^a}{d\tau}} = \Omega^a{}_b s^b,\qquad \hbox{or}\qquad
{\frac {d{\bm s}}{d\tau}} = {\bm \Omega}\times{\bm s}.\label{ds1}
\end{equation}
Here ${\bm s} = \{s^a\}$ and ${\bm\Omega} = \left\{-\,{\frac 12}\epsilon^{abc}\Omega_{bc}\right\}$. As a result, we find the precession angular velocity in terms of the magnetic and electric projections of the spin transport matrix (\ref{NQ}):
\begin{equation}
\bm{\Omega} = \gamma\,\Bigl\{\bm{Q} - \widehat{\bm v}\,Q^0 + {\frac \gamma{\gamma + 1}}
\,{\frac {1}{c^2}}\left[\widehat{\bm v}\,(\widehat{\bm v}\cdot\bm{Q})  - \widehat{v}{}^2\bm{Q}
\right]\Bigr\} + {\frac {\gamma}{\gamma + 1}}\,{\frac {\bm{\mathcal F}\times\widehat{\bm v}}{c^2}}.\label{OG0}
\end{equation}
Quite remarkably, it turns out that $\bm{\Omega}$ does not depend on the ``electric'' part $N^\alpha$, and moreover it actually depends only on the transversal part of the ``magnetic'' vector, namely ${}^\bot\!Q^\alpha = \rho^\alpha_\beta Q^\beta$. Explicitly in components we have
\begin{equation}
{}^\bot\!Q^0 = {\frac {\gamma^2}{c^2}}\left[\widehat{\bm v}\cdot\bm{Q} - \widehat{v}{}^2Q^0\right],
\qquad {}^\bot\!Q^a = Q^a + {\frac {\gamma^2}{c^2}}\left[\widehat{\bm v}\cdot\bm{Q} - c^2Q^0\right]
\widehat{v}^a,\label{QQ}
\end{equation}
and one can check the useful relations
\begin{equation}
{}^\bot\!\bm{Q} - {}^\bot\!Q^0\,\widehat{\bm v} = \bm{Q} - Q^0\,\widehat{\bm v},\qquad
({}^\bot\!\bm{Q}\cdot\widehat{\bm v})\,\widehat{\bm v}  - {}^\bot\!\bm{Q}\,\widehat{v}{}^2
= (\bm{Q}\cdot\widehat{\bm v})\,\widehat{\bm v}  - \bm{Q}\,\widehat{v}{}^2.\label{vQ}
\end{equation}
From (\ref{QQ}) we then derive
\begin{equation}
{}^\bot\!Q^0 = {\frac 1{c^2}}\,{}^\bot\!\bm{Q}\cdot\widehat{\bm v},\label{Qbot} 
\end{equation}
and recast (\ref{OG0}) into a final form 
\begin{equation}
\bm{\Omega} = {}^\bot\!\bm{Q} - {\frac {\gamma}{\gamma + 1}}\,{\frac {({}^\bot\!\bm{Q}\cdot\widehat{\bm v})\,\widehat{\bm v}}{c^2}} - {\frac {\gamma}{\gamma + 1}}\,{\frac {\widehat{\bm v}\times\bm{\mathcal F}}{c^2}}.\label{OG}
\end{equation}
The new general equations for the spin precession (\ref{OG}) and (\ref{ds1}) are valid for a particle moving in arbitrary external fields. The actual dynamics of the physical spin depends on the forces acting on the particle and on the spin transport law.

\section{Spin in the gravitational and electromagnetic fields}

In order to discuss the spin dynamics in external gravitational (inertial) and electromagnetic fields, we recall that the gravitational field is described by an arbitrary coframe and a local Lorentz connection $e^\alpha_i$ and $\Gamma_i{}^{\alpha\beta} = -\Gamma_i{}^{\beta\alpha}$; the spacetime metric is then constructed as $g_{ij} = e^\alpha_i e^\beta_j g_{\alpha\beta}$. In the Schwinger gauge, the coframe is parametrized as
\begin{equation}\label{coframe}
e_i^{\,\widehat{0}} = V\,\delta^{\,0}_i,\qquad e_i^{\widehat{a}} =
W^{\widehat a}{}_b\left(\delta^b_i - cK^b\,\delta^{\,0}_i\right),\qquad a,b=1,2,3,
\end{equation}
in terms of the functions $V(t, \bm{x}), \bm{K}(t, \bm{x}), W^{\widehat a}{}_b(t, \bm{x})$ which depend arbitrarily on the local coordinates $\left\{t, \bm{x}\right\}$ on the spacetime manifold.

The classical theory of spin was developed soon after the concept of spin was proposed in particle physics (see \cite{corben} for introduction and history). This theory underlies the analysis of the dynamics of polarized particles in accelerators and storage rings \cite{derb}.
Neglecting second-order spin effects \cite{chicone}, the dynamical equations for a charged spinning particle (in general, with magnetic and electric dipole moments) in external electromagnetic and gravitational fields are written as
\begin{eqnarray}
{\frac {DU^\alpha}{d\tau}} &=& -\,{\frac em}\,F^\alpha{}_\beta\,U^\beta,\label{eomU}\\
{\frac {DS^\alpha}{d\tau}} &=& - \,{\frac em}\,F^\alpha{}_\beta S^\beta 
-\,{\frac 2\hbar}\,{}^\bot\!M^\alpha{}_\beta S^\beta .\label{eomS}
\end{eqnarray}
Here we follow the notations and conventions introduced in the previous sections. 

Comparing the system (\ref{eomU})-(\ref{eomS}) with (\ref{dotUS}), we find explicitly the generalized force and the spin transport matrix:
\begin{eqnarray}\label{FePe}
{\mathcal F}_\alpha = \phi_{\alpha\beta}U^\beta,\qquad 
\phi_{\alpha\beta} = \Phi_{\alpha\beta} + U^i\Gamma_{i\alpha\beta}.
\end{eqnarray}
The latter can be conveniently recast into a more compact form $\phi_{\alpha\beta} = -\,{\frac {e}{m}}\,\dF_{\alpha\beta}$ if we introduce the combined (or ``artificial electromagnetic'') field
\begin{equation}
\dF_{\alpha\beta} = F_{\alpha\beta} + {\frac {2m}{e\hbar}}\,{}^\bot\!M_{\alpha\beta}
- {\frac {m}{e}}\,U^i\Gamma_{i\alpha\beta}.\label{Fc}
\end{equation}
As usual, we identify the components of the tensor (\ref{Mab}) with the polarization and magnetization 3-vectors $c\bm{\mathcal P}{}_a = \{M_{\widehat{0}\widehat{1}}, M_{\widehat{0}\widehat{2}}, M_{\widehat{0}\widehat{3}} \}, \bm{\mathcal M}{}^a = \{ M_{\widehat{2}\widehat{3}}, M_{\widehat{3}\widehat{1}}, M_{\widehat{1}\widehat{2}} \}$, respectively.

As a result, we obtain the components of the projected transversal tensor: 
\begin{eqnarray}
{}^\bot\!M_{0a} &=& {\frac {\gamma^2}{c}}\left\{-\,v^2{\mathcal P}_a + \widehat{v}_a(\widehat{\bm v}
\cdot\bm{\mathcal P}) + c[\widehat{\bm v}\times\bm{\mathcal M}]_a\right\},\label{Mb0}\\
{}^\bot\!M_{ab} &=& \epsilon_{abc}\,{\frac {\gamma^2}{c^2}}\left\{c^2{\mathcal M}^c - \widehat{v}^c
(\widehat{\bm v}\cdot\bm{\mathcal M}) + c[\widehat{\bm v}\times\bm{\mathcal P}]^c\right\}.\label{Mba}
\end{eqnarray}
Recalling that the components of the true electric and magnetic fields are introduced as $\bm{\mathfrak{E}}{}_a = \{F_{\widehat{1}\widehat{0}}, F_{\widehat{2}\widehat{0}}, F_{\widehat{3}\widehat{0}} \}, \bm{\mathfrak{B}}{}^a = \{F_{\widehat{2}\widehat{3}}, F_{\widehat{3}\widehat{1}}, F_{\widehat{1}\widehat{2}} \}$, by analogy we define the effective ``electric'' and ``magnetic'' fields
\begin{equation}\label{EBeff}
\bm{\mathfrak{E}}{}_a^{\rm eff} = \{ \dF_{\widehat{1}\widehat{0}}, \dF_{\widehat{2}\widehat{0}},
\dF_{\widehat{3}\widehat{0}} \},\qquad\bm{\mathfrak{B}}{}^a_{\rm eff} = \{\dF_{\widehat{2}\widehat{3}},
\dF_{\widehat{3}\widehat{1}}, \dF_{\widehat{1}\widehat{2}} \}.
\end{equation}
Making use of (\ref{Fc}), (\ref{Mb0}) and (\ref{Mba}), we have explicitly
\begin{eqnarray}
\bm{\mathfrak{E}}_{\rm eff} &=& \bm{\mathfrak{E}} - {\frac {2m}{e\hbar}}\,\gamma^2\,
\widehat{\bm{v}}\times\bm{\Delta} + {\frac me}\,\bm{\mathcal E},\label{EEE}\\
\bm{\mathfrak{B}}_{\rm eff} &=& \bm{\mathfrak{B}} + {\frac {2m}{e\hbar}}\,\gamma^2
\Bigl[\bm{\Delta} - {\frac 1{c^2}}\,\widehat{\bm{v}}\,(\widehat{\bm{v}}\cdot\bm{\Delta})
\Bigr] + {\frac me}\,\bm{\mathcal B}.\label{BBB}
\end{eqnarray}
Here the generalized polarization current
\begin{equation}\label{Delta}
\bm{\Delta} = \bm{\mathcal M} + {\frac 1c}\,\widehat{\bm{v}}\times\bm{\mathcal P}
\end{equation}
accounts for the electromagnetic nonminimal coupling effects, whereas the gravitoelectric and gravitomagnetic fields
\begin{equation}\label{EBTN}
\bm{\mathcal E}_a = -\,U^i\Gamma_{i\hat{a}\hat{0}},\qquad
\bm{\mathcal B}^a = -\,{\frac 12}\,\epsilon^{abc}\,U^i\Gamma_{i\hat{b}\hat{c}}
\end{equation}
encompass general-relativistic gravitational and/or inertial contributions.

Applying the projection method, we decompose the artificial electromagnetic field (\ref{Fc}) 
\begin{equation}
\dF_{\alpha\beta} = {\frac 1{c^2}}\left(\dE_\alpha U_\beta - \dE_\beta U_\alpha
+ c\eta_{\alpha\beta\mu\nu}U^\mu\dB^\nu\right)\label{FEB1}
\end{equation}
in terms of the ``electric'' and ``magnetic'' 4-vectors 
\begin{equation}
\dE_\alpha := \dF_{\alpha\beta}U^\beta,\qquad \dB^\alpha := {\frac 1{2c}}
\eta^{\alpha\mu\nu\beta}\dF_{\mu\nu}U_\beta.\label{EBc}
\end{equation}
Comparing (\ref{FEB1}) and (\ref{EBc}) with (\ref{NQ}) and (\ref{PNQ}), we identify 
\begin{eqnarray}
N_\alpha = -\,{\frac em}\,\dE_\alpha,\qquad
Q^\alpha = -\,{\frac em}\,\dB^\alpha,\label{NEQB}  
\end{eqnarray}
and consequently we derive explicitly the components 
\begin{eqnarray}
N^0 = \gamma\,{\frac {e\widehat{\bm v}\cdot\bm{\mathfrak{E}}{}_{\rm eff}}{mc^2}}, &\qquad&
\bm{N} = \gamma\,{\frac em}\left(\bm{\mathfrak{E}}{}_{\rm eff} + \widehat{\bm v}\times
\bm{\mathfrak{B}}{}_{\rm eff}\right),\label{Neff}\\
Q^0 = -\,\gamma\,{\frac {e\widehat{\bm v}\cdot\bm{\mathfrak{B}}{}_{\rm eff}}{mc^2}}, &\qquad&
\bm{Q} = -\,\gamma\,{\frac em}\Bigl(\bm{\mathfrak{B}}{}_{\rm eff} - {\frac 1{c^2}}
\widehat{\bm v}\times\bm{\mathfrak{E}}{}_{\rm eff}\Bigr).\label{Qeff}
\end{eqnarray}
By construction, ${}^\bot\!\bm{Q} = \bm{Q}$, and moreover we have $\bm{\mathcal F} = \bm{N}$. Inserting (\ref{Neff}) and (\ref{Qeff}) into (\ref{OG}), we obtain the spin precession angular velocity 
\begin{equation}\label{Otot}
\bm{\Omega} = {\frac em}\left( -\,\bm{\mathfrak{B}}{}_{\rm eff} + {\frac {\gamma}
{\gamma + 1}}\,{\frac {\widehat{\bm{v}}\times\bm{\mathfrak{E}}{}_{\rm eff}}{c^2}}\right)
\end{equation}
as a function of external fields which enter via the effective variables (\ref{EEE})-(\ref{BBB}).

\section{Quantum spinning particle}

The quantum dynamics of a fermion particle with spin ${\frac 12}$, electric charge $e$ and the rest mass $m$ is described by the relativistic Dirac theory. The 4-spinor field $\psi$ satisfies the covariant wave equation
\begin{eqnarray}\label{LD}
i\hbar\gamma^\alpha e_\alpha^i \left(\partial _i\psi - {\frac {ie}{\hbar}}\,A_i\psi
+ {\frac i4}\Gamma_i{}^{\beta\gamma}\sigma_{\beta\gamma}\psi\right) - mc\,\psi 
+ {\frac {1}{2c}}M_{\alpha\beta}\,\sigma^{\alpha\beta}\psi = 0,
\end{eqnarray}
where $\sigma_{\alpha\beta} = i\gamma_{[\alpha} \gamma_{\beta]}$. The first two terms in (\ref{LD}) describe the minimal coupling of the spinor field to the electromagnetic $A_i = (-\,\Phi, \bm{A})$ and the gravitational $e^\alpha_i$ and $\Gamma_i{}^{\alpha\beta}$ fields. In addition, we assume a possible non-minimal interaction due to particle's nontrivial magnetic and electric dipole moments, which is described by the last Pauli-type term in (\ref{LD}), where the generalized polarization tensor is given by (\ref{Mab}). 

To reveal the physical contents of the relativistic quantum theory, we recast the Dirac equation (\ref{LD}) into a Schr\"odinger form $i\hbar {\frac {\partial\psi}{\partial t}} = {\mathcal H}\psi$ and go to the Foldy-Wouthuysen (FW) representation by means of the unitary transformation $\psi_{FW} = U\psi$ and ${\mathcal H}_{FW} = U{\mathcal H}U^{-1} - i\hbar U\partial_tU^{-1}$. The resulting FW Hamiltonian in the semiclassical approximation then reads
\begin{equation}
{\mathcal H}_{FW} = \beta mc^2V\gamma + e\Phi + {\frac c2}\left(\bm{K}\cdot\bm{\pi}
+ \bm{\pi}\cdot\bm{K}\right) + {\frac \hbar 2}\bm{\Sigma}\cdot\bm{\Omega}.\label{Hamlt}
\end{equation}
Here $\bm\pi=-i\hbar\bm{\nabla} - e\bm A$ is the kinetic momentum operator and $\bm\Sigma$ is the spin matrix; as before, the Schwinger gauge is used for the parametrization of the coframe by means of (\ref{coframe}). The crucial observation is that the momentum operator $\bm\pi$ is related to the velocity operator $\widehat{\bm v}$ via $\beta\,W^b{}_{\widehat a}\pi_b = m\gamma\widehat{v}{}_a$. As a result, the spin precession angular velocity operator $\bm{\Omega}$, expressed in terms of $\bm\pi$, has exactly the form (\ref{Otot}). Therefore, it is satisfactory to see that the classical and quantum spin dynamics turn out to be fully consistent.

\section{Discussion and conclusion}

A unified approach to the study of classical and quantum spin in arbitrary external fields of different physical nature is developed here on the basis of projections of the physical quantities, specified by the 4-velocity of a particle. The method is applied to the motion of spin under the action of the electromagnetic and gravitational fields. It is worthwhile to notice that one can consistently generalize the formalism so that to take into account possible post-Riemannian deviations of the spacetime geometry \cite{ost4}. The results obtained are of direct operational relevance and they form the basis for many physical observations and experiments in high-energy physics and astrophysics. Important applications range from the neutrino physics in matter \cite{max} to study of polarized beams in accelerators and storage rings \cite{ost5} to the development of new methods of gravitational wave detection \cite{ost6}. Of particular interest is to use the new formalism for the analysis of the chiral phenomena in the heavy-ion collisions \cite{pro1,pro2,pro3}. 

I thank the organizers of DSPIN2019 for the hospitality and support. This work was partially supported by the Russian Foundation for Basic Research (Grant No. 18-02-40056-mega).

\end{document}